\documentclass[aps,prl,twocolumn,amsmath,amssymb,showpacs, superscriptaddress,notitlepage,longbibliography,10pt]{revtex4-1}

\usepackage[colorlinks=true,linkcolor=black,anchorcolor=red,citecolor=black, urlcolor=black]{hyperref}
\usepackage{bm}
\usepackage{graphicx}
\usepackage{color}
\usepackage{multirow}
\usepackage{extarrows}
\usepackage{graphicx}
\usepackage{dcolumn}
\usepackage{bm}
\usepackage{xcolor, color}
\usepackage{cases}
\usepackage{bbding}

\usepackage{xr}

\usepackage{makecell}

\begin{document}
\title{Tunable Quantum-Metric-Driven Nonlinear Valley Hall Effect in \(\mathcal{P,T}\)-symmetric Sliding Antiferroelectric MoS$_2$ } 

\author{Wen-Xin Jiang}
\thanks{These authors contributed equally to this work}
\affiliation{State Key Laboratory of Quantum Functional Materials, Department of Physics, and Guangdong Basic Research Center of Excellence for Quantum Science, Southern University of Science and Technology, Shenzhen, 518055, China}

\author{Zhen-Hao Gong}
\thanks{These authors contributed equally to this work}
\affiliation{State Key Laboratory of Quantum Functional Materials, Department of Physics, and Guangdong Basic Research Center of Excellence for Quantum Science, Southern University of Science and Technology, Shenzhen, 518055, China}

\author{Zhigang Gui}
\affiliation{Quantum Science Center of Guangdong-Hong Kong-Macao Greater Bay Area (Guangdong), Shenzhen 518045, China}

\author{M. Umar Farooq}
\email{farooqmu@sustech.edu.cn}
\affiliation{State Key Laboratory of Quantum Functional Materials, Department of Physics, and Guangdong Basic Research Center of Excellence for Quantum Science, Southern University of Science and Technology, Shenzhen, 518055, China}

\author{Li Huang}
\email{huangl@sustech.edu.cn}
\affiliation{State Key Laboratory of Quantum Functional Materials, Department of Physics, and Guangdong Basic Research Center of Excellence for Quantum Science, Southern University of Science and Technology, Shenzhen, 518055, China}
\affiliation{Quantum Science Center of Guangdong-Hong Kong-Macao Greater Bay Area (Guangdong), Shenzhen 518045, China}

\date{\today}

\begin{abstract}
Quantum metric, the real part of quantum geometric tensor, plays an important role in fundamental transport and optical responses of quantum materials. While net quantum-metric effects are typically realized by breaking both time-reversal ($\mathcal T$) and inversion ($\mathcal P$) symmetries, valley-contrast mechanisms enable the probing of quantum geometry in centrosymmetric, nonmagnetic systems, revealing physical signatures beyond Berry curvature. Here we design a four-layer MoS$_2$ platform that combines sliding antiferroelectric order with strain-induced $C_{3z}$ breaking to realize a quantum-metric-driven nonlinear valley Hall effect (NVHE) under preserved $\mathcal P$ and $\mathcal T$ symmetries. Reversible interlayer sliding (ABBA$\leftrightarrow$BAAB) provides non-volatile electrical control of the valley response. First-principles calculations reveal an outer-layer-dominated NVHE with tunable sign and magnitude. This symmetry-preserving, sliding-tunable strategy activates hidden quantum-geometric phenomena in realistic 2D materials and opens a clean pathway for tunable valleytronics and quantum-information devices.

\end{abstract}

\maketitle

\begingroup
\footnotesize
\raggedright
\noindent
\copyright\ 2026 American Physical Society.
This is the accepted manuscript of the following article:
Wen-Xin Jiang, Zhen-Hao Gong, Zhigang Gui,
M. Umar Farooq, and Li Huang,
``Tunable quantum metric driven nonlinear valley Hall effect
in the $\mathcal{PT}$-symmetric sliding antiferroelectric
MoS$_2$'',
Phys. Rev. B \textbf{113}, 235116 (2026).
The final published version is available from
\url{https://doi.org/10.1103/p9sb-k9fj}.
This manuscript version is posted with permission
for non-commercial scholarly use.
\par
\endgroup


\section{\label{sec:level1}I. Introduction}

\begin{figure*}
\includegraphics[scale=0.52]{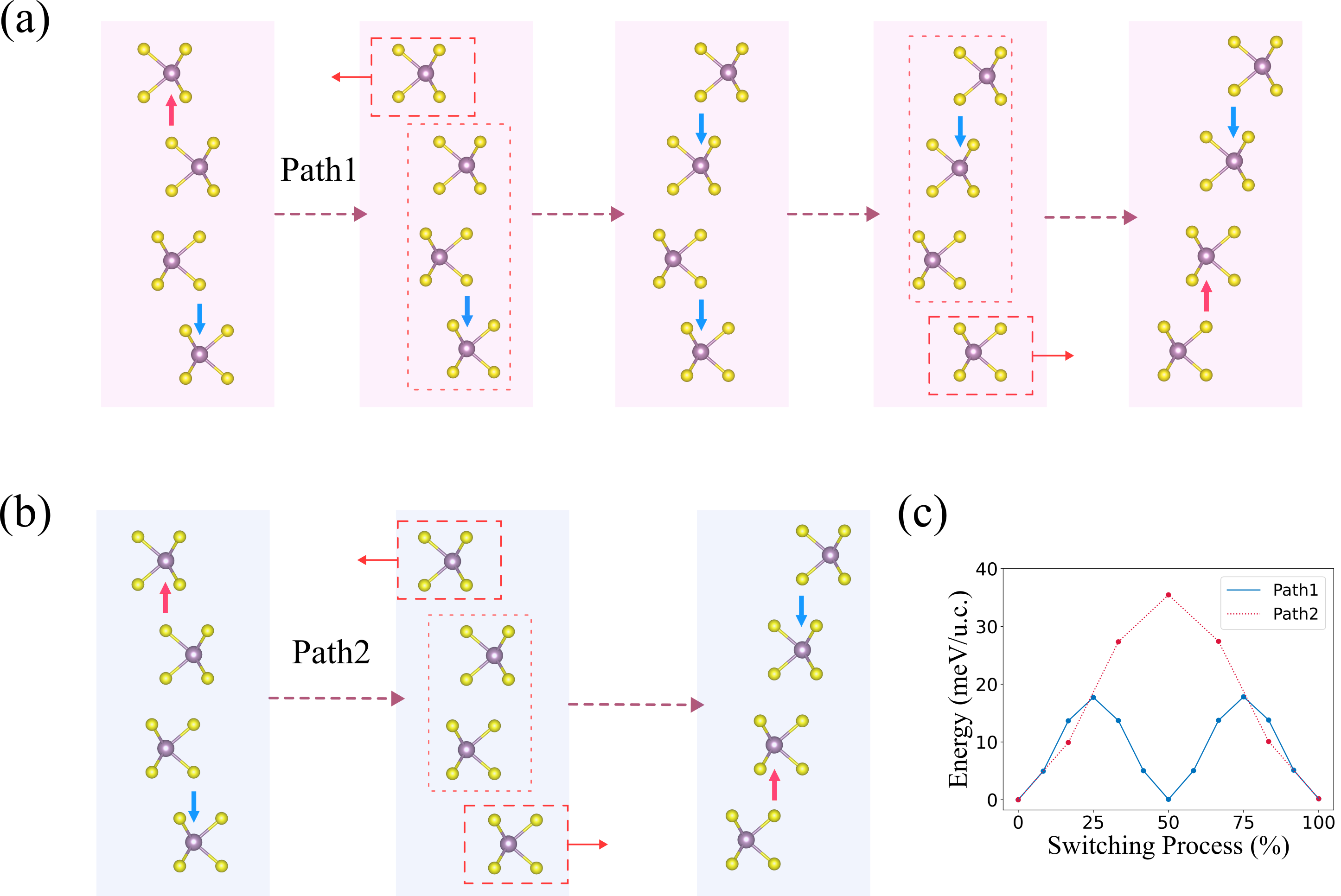}
\caption{Sliding pathways and energetics of antiferroelectric (AFE) switching in centrosymmetric four-layer MoS$_2$. (a,b) Schematic diagrams of two possible AFE switching pathways between ABBA and BAAB stackings. (a) 
Path 1: sequential sliding of the topmost layer followed by the bottommost layer. (b) Path 2: simultaneous sliding 
of both outer layers. In both pathways, the inner bilayer maintains the 2H stacking throughout, ensuring that \(\mathcal P\) and \(\mathcal T\) are fully restored in the final states. Horizontal red arrows indicate the in-plane sliding direction of the layer(s) within the corresponding red dashed boxes. Vertical blue (downward) and red (upward) arrows represent the local electric polarization orientation in each bilayer. (c) Total energy profiles (per unit cell) along the switching coordinate for both pathways. The lower energy barrier along Path 1 ($\sim$17.7 meV/unit cell) indicates that it is the more favorable route for reversible, electrically accessible AFE switching.}\label{sliding}
\end{figure*}

The family of Hall effects provides sensitive probes of band geometry and symmetry in quantum materials, underpinning transport metrology and low-dissipation electronics. In time-reversal–symmetric systems, the Berry curvature satisfies $\Omega_n(\mathbf{k})=-\Omega_n(-\mathbf{k})$, enforcing a vanishing net linear Hall conductivity. The valley Hall effect (VHE) leverages the opposite Berry curvature at the $K$ and $K'$ valleys, where carriers acquire transverse velocities of opposite sign, yielding zero net transverse charge current but finite valley Hall current~\cite{Di_Xiao-2007-PRL-valley-grahene}. This valley-contrasting transport has been observed via valley-selective optical circular dichroism and nonlocal electrical measurements, where an in-plane bias drives valley-polarized currents to opposite sample edges~\cite{Huangling_Zeng-2012-Nat_Nanotech-Valley-MoS2-experiment-optical,Mengqiao_Sui-2015-Nat_Phys-VHE-2L-graphen-experiment,Y_Shimazaki-2015-Nat_Phys-VHE-2L-graphen-experiment}. Despite these advances, the conceptual framework of the VHE remains predominantly anchored to the Berry curvature. A more complete geometric picture is encoded in the quantum-geometric tensor, whose imaginary part gives the Berry curvature, while its real part, the quantum metric, quantifies the distance between neighboring Bloch states in Hilbert space and their evolution in momentum space~\cite{JP_Provost-1980-Riemannian,Resta-2011-geometrical_theory,Rahul_Roy-2014-Phys_Rev_B-band_geometry,Peotta_Sebastiano-2015-Nat_Commun-superfluidity_topology,Torma-2022-Nat_Rev_Phys-superconductivity_superfluidity_quantum_geometry,Torma-2023-PRL-Essay_QMD,Hetenyi-2023-PRA-geometry_qauntum_state}. Although the roles of Berry curvature and its dipole (BCD) in linear and nonlinear transport are well-established~\cite{Liang_Fu-2015-PRL-BCD}, the transport consequences of the quantum metric and its dipole (QMD) have only recently emerged as a frontier, offering a new paradigm for understanding and engineering electronic responses beyond the adiabatic approximation.

The QMD has been shown to drive nonlinear Hall responses in magnetic materials such as CuMnAs~\cite{Chong_Wang-2021-PRL-CuMnAs}, Mn$_2$Au~\cite{Huiying_Liu-2021-PRL-Mn2Au}, and even-layer MnBi$_2$Te$_4$~\cite{Anyuan_Gao-2023-Science-QMD-MnBiTe-experiment,Naizhou_Wang-2023-Nature-QMD-experiment,Gong-2024-QMD-theory}. This behavior is fundamentally governed by symmetry: the QMD is odd under both inversion ($\mathcal P$) and time reversal ($\mathcal T$), but even under their combined operation $\mathcal{PT}$. Consequently, a net QMD can emerge in systems that break $\mathcal P$ and $\mathcal T$ individually while preserving $\mathcal{PT}$. In real magnetic systems, however, parasitic signals arising from unintentional symmetry lowering (e.g., due to strain and inhomogeneity), current-driven domain reconfiguration~\cite{Reimers-2023-NC-current-domain-Mn2Au} and thermoelectric backgrounds induced by Joule heating and temperature gradients~\cite{Sheng_Peng-2022-thermal_contribution} can produce extrinsic transverse signals, often obscuring the intrinsic geometric response, complicating the isolation of pure QMD contributions. These challenges motivate the pursuit of an alternative route: accessing the QMD in nonmagnetic, centrosymmetric systems, where strict symmetry control eliminates magnetic noise and provides a cleaner platform for probing quantum-geometric transport.

To isolate the intrinsic quantum-metric response, we target a nonmagnetic, inversion-symmetric regime, where Berry curvature and its dipole (BCD) are strictly forbidden by symmetry, and magnetic noise is absent. This creates an ideal platform where any surviving nonlinear valley response provides a clean window into the quantum metric, even though the Brillouin zone (BZ)-integrated QMD vanishes. 
{\color{black}Previous theoretical studies have shown that,} in such centrosymmetric systems, a quantum metric driven nonlinear valley Hall effect (NVHE) can be realized by selectively reading out the valley contrasting QMD 
distributions~\cite{Kamal_Das-2024-PRL-NVHE,Zhichao_Zhou-2025-PRB-NVHE-2L-MoS2}. {\color{black} Correspondingly, the experimental identification of the nonlinear valley Hall effect in centrosymmetric systems remains an interesting direction for further exploration~\cite{Cao_Jin-2025-Nonlocal-NVHE}.}
Two-dimensional transition metal dichalcogenides (TMDs) provide an ideal platform for this purpose, given their tunable band structures and symmetry-governed transport properties. However, in the widely studied 2H-phase TMDs, the intrinsic $C_{3z}$ rotational symmetry forbids a finite QMD. Applying uniaxial strain breaks $C_{3z}$ while preserving both nonmagnetic character and global inversion symmetry, thereby lifting this restriction and enabling nonzero, opposite sign QMD at the $K$ and $K'$ valleys. This symmetry-guided strategy has been validated in bilayer 2H-MoS$_2$, where strain-induced $C_{3z}$ breaking generates QMD-driven NVHE signals without destroying centrosymmetry~\cite{Zhichao_Zhou-2025-PRB-NVHE-2L-MoS2}. While such bilayer systems have successfully demonstrated strain-enabled, QMD-driven NVHE in centrosymmetric and nonmagnetic platforms, they generally lack additional knobs for reversible control of the nonlinear response. 

In this work, we propose a four-layer MoS$_2$ architecture that integrates sliding antiferroelectric (AFE) order as an active tuning knob for QMD-driven NVHE. In this design, out-of-plane layer dipoles are arranged in an AFE pattern, canceling the net polarization and preserving global inversion symmetry. The system supports two reversible stacking configurations, ABBA and BAAB, linked by interlayer sliding, each exhibiting an opposite interlayer polarization pattern. This enables non-volatile on/off and sign control of valley-selective NVHE without breaking centrosymmetry or introducing magnetism. First-principles calculations further reveal a pronounced layer-valley locking effect: in the ABBA configuration, conduction states near the $K/K'$ valleys are predominantly localized on the outer layers, whereas in BAAB they shift to the inner layers. By combining uniaxial strain to break $C_{3z}$ symmetry with sliding-controlled AFE switching, our platform achieves centrosymmetry-preserving, reversible electrical tunability of QMD-driven nonlinear responses. This dual-control strategy establishes a versatile materials platform for tunable valleytronic logic and quantum-information processing devices.

\begin{figure*}
\includegraphics[scale=0.4]{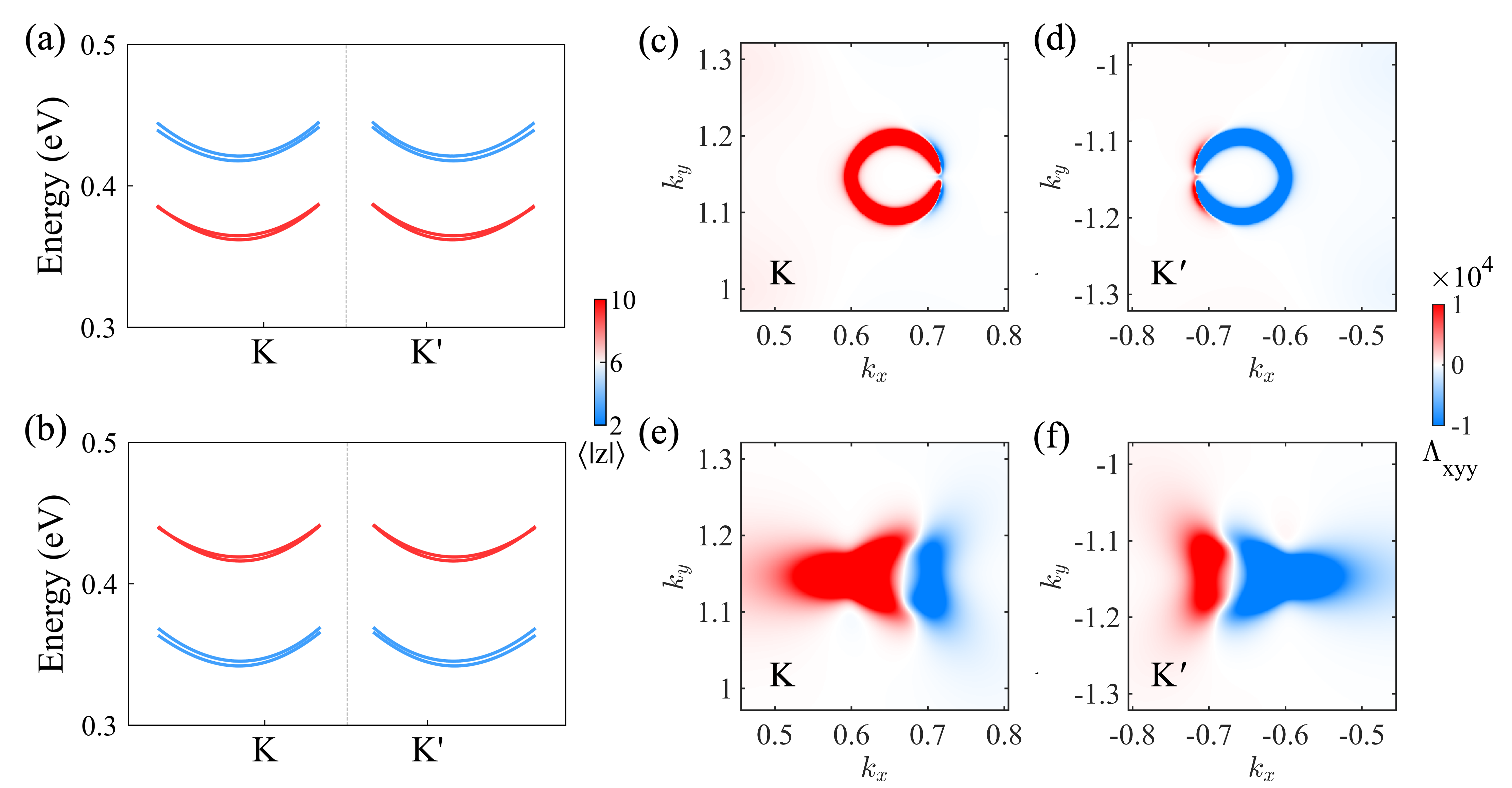}
\caption{
Strain-engineered band structures and layer-resolved QMD in four-layer MoS$_2$. (a,b) Conduction bands near the \(K\) and \(K'\) valleys for ABBA- and BAAB-stacked configurations under 5\% uniaxial strain along the zigzag direction. {\color{black}The bands are colored by the expectation value \(\langle |z| \rangle\), obtained from the atomic-projected weights, which provides a continuous measure of the layer distribution of the Bloch states: smaller values correspond to stronger inner-layer character, whereas larger values indicate stronger outer-layer character. The complete projected band structure is provided in Fig.~S2} \cite{supp}. While the bands remain doubly degenerate due to preserved \(\mathcal{P}\) and \(\mathcal{T}\) symmetries, the broken \(C_{3z}\) symmetry activates valley-contrasting QMD. (c,d) Layer-resolved QMD distribution near the conduction band minima for ABBA stacking, revealing dominant contributions from outer layers. (e,f) Corresponding QMD distributions for BAAB stacking, showing enhanced weight on inner layers. These distinct layer–valley–QMD correlations demonstrate reversible, stacking-controlled tunability of the nonlinear valley Hall response.
}
\label{QMD}
\end{figure*}

\section{\label{sec:level2}II. Computational methodology}

Our first-principles calculations are performed within the density functional theory (DFT) framework using the projected augmented wave method, as implemented in the Vienna ab initio Simulation Package (VASP)~\cite{Blochl-1994-PAW,kresse-1994-GGA,kresse-1996-efficient-iterative-schemes}. The exchange-correlation functional is treated using the generalized gradient approximation (GGA) with the Perdew-Burke-Ernzerhof (PBE) functional \cite{perdew-1996-PBE,kresse-1999-PAW}. The Kohn-Sham orbitals are expanded in a plane-wave basis set with an energy cutoff of 500 eV. Structures are fully relaxed until the atomic forces are below \(10^{-2}\) eV/\AA, and the energy differences between consecutive self-consistent steps are less than \(10^{-5}\) eV. The BZ is sampled using the Monkhorst-Pack scheme with a grid of 12 × 12 × 1 k-points centered at the \(\Gamma\) point. A vacuum space of 20 \AA~is included in the non-periodic direction to prevent interactions between adjacent images. The vdW interactions between the layers are described by the DFT-D3 method \cite{Grimme-2011-DFT_D3-JCC}. Convergence tests are performed with respect to the energy cutoff and $k$-point sampling. 

Subsequently, the Bloch wave functions obtained from DFT are projected onto maximally localized Wannier functions (MLWFs) \cite{Wannier90} to construct a tight-binding model comprising 88 bands (44 orbital channels with two spin components) derived from Mo $d$ and S $p$ orbitals. The resulting Wannier-based tight-binding model is used to compute the QMD and NVHE with high spectral resolution near the Fermi level.

\begin{table}[h]
	\caption{\label{tab:hall_summary}Symmetry-allowed quantum-geometry--related Hall-type responses in monolayer (1L), bilayer (2L), and four-layer (4L) MoS$_2$ systems.
Key symmetries ($C_{3z}$, $\mathcal{P}$, $\mathcal{T}$) are marked as present (\Checkmark) or broken (\XSolidBrush).
Quantum-geometry--related Hall-type responses considered here include those induced by the Berry curvature (BC), Berry-curvature dipole (BCD), and quantum-metric dipole (QMD) \cite{Di_Xiao-2012-PRL-MoS2-valley, xiong-2025-JPCM-MoS2_BCD, Zhichao_Zhou-2025-PRB-NVHE-2L-MoS2, Huiying_Liu-2021-PRL-Mn2Au, ZZ_Du-2021-NC-NHE}. 
Superscript $\text{v}$ denotes responses accessible via valley-resolved measurements.
}
\begin{ruledtabular}
\begin{tabular}{lccccc}
System & \(C_{3z}\) & \(\mathcal{P}\) & \(\mathcal{T}\) & \(\mathcal{PT}\) & Hall Signal \\
\hline
1L / 2L-FE (3R)          & \Checkmark & \XSolidBrush & \Checkmark & \XSolidBrush & BC$^\text{v}$ \\
1L / 2L-FE (3R) + strain         & \XSolidBrush & \XSolidBrush & \Checkmark & \XSolidBrush & \makecell{BC$^\text{v}$, \\ BCD, QMD$^\text{v}$} \\
2L (2H) / 4L-AFE           & \Checkmark & \Checkmark & \Checkmark & \Checkmark & None \\
2L (2H) / 4L-AFE + strain          & \XSolidBrush & \Checkmark & \Checkmark & \Checkmark & QMD$^\text{v}$\\
\end{tabular}
\end{ruledtabular}
\end{table}

\begin{figure*}
\includegraphics[scale=0.45]{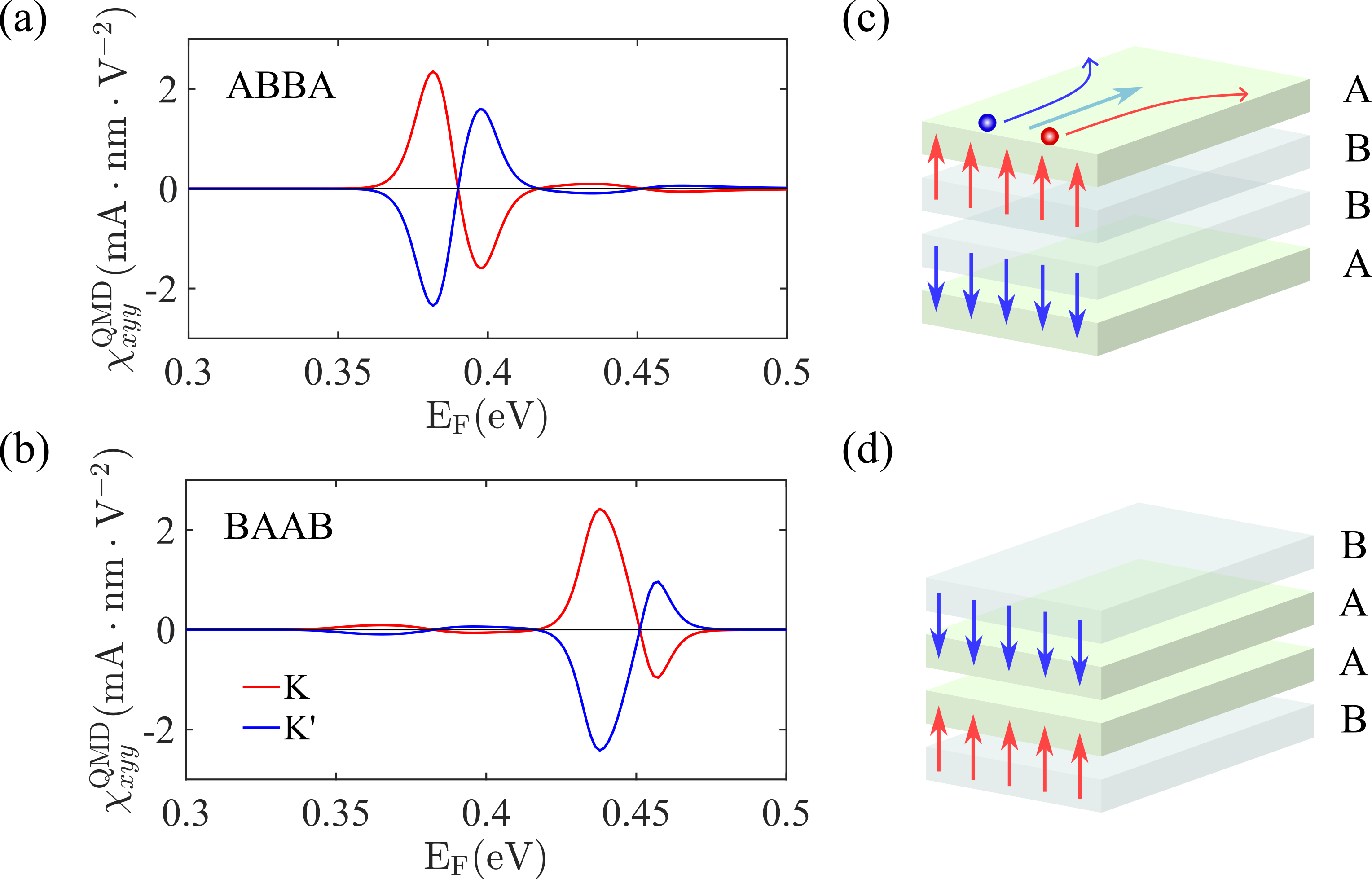}
\caption{
Layer-resolved nonlinear valley Hall conductivity (NVHC) and {\color{black}schematic illustrations}. (a, b) NVHC for ABBA- and BAAB-stacked four-layer MoS$_2$ under 5\% uniaxial strain at 40 K, showing a pronounced stacking-dependent and layer-selective response dominated by quantum metric effects. Results for 1\%–4\% strain are provided in Fig.~S5 \cite{supp}. {\color{black}(c, d) Schematic illustrations of the layer-resolved NVHE in the two opposite AFE stackings. In each panel, the green-shaded layer denotes the layer making the dominant contribution to the transport response, while the transparent layer indicates a comparatively weaker contribution. (c) highlights the valley-contrasting transverse deflection of carriers in the stacking configuration with dominant outer-layer contribution to the response, whereas (d) illustrates the corresponding layer-resolved NVHE pattern for the opposite stacking, where the dominant contribution is shifted toward the inner layers.}}
\label{QMD}
\end{figure*}

\section{III. Results and discussion}

We begin by systematically analyzing the evolution of structure and symmetry from monolayer to bilayer, and finally to our proposed four-layer architecture. Monolayer MoS$_2$ adopts a trigonal prismatic coordination, with each Mo atom hexagonally coordinated by six S atoms in an S--Mo--S trilayer structure (Fig. 1a). The monolayer lacks inversion symmetry (point group $D_{3h}$), which permits finite Berry curvature \cite{Di_Xiao-2012-PRL-MoS2-valley} and second-order conductivity \cite{You_Jhih-Shih-2018-PRB-MoS2-BCP}. However, time-reversal symmetry enforces $\Omega_n(\mathbf{k})=-\Omega_n(-\mathbf{k})$, causing the linear Hall conductivity to vanish upon BZ integration, thereby making valley-resolved analysis essential. More critically, the threefold rotational symmetry ($C_{3z}$) inherent to the monolayer imposes a stringent constraint on intrinsic second-order Hall responses: it forbids both the Berry-curvature dipole (BCD) and the quantum-metric dipole (QMD). This prohibition is not merely a consequence of BZ cancellation; it holds rigorously even within a single valley (little group $C_{3h}$). Breaking $C_{3z}$ via uniaxial strain reduces the symmetry (e.g., to $C_s$) and lifts this restriction, enabling non-zero, symmetry-allowed Hall-type second-order responses where both BCD and QMD mechanisms can, in principle, contribute \cite{Huiying_Liu-2021-PRL-Mn2Au,ZZ_Du-2021-NC-NHE}.

The bilayer case highlights the crucial role of inversion symmetry. In the 2H-stacked bilayer, inversion symmetry is restored, changing the point group from $D_{3h}$ (monolayer) to $D_{3d}$. This symmetry enforces $\Omega_n(\mathbf{k})=\Omega_n(-\mathbf{k})$, which, together with time-reversal symmetry, mandates a strictly vanishing Berry curvature across the entire BZ. Consequently, both linear Hall currents induced by Berry curvature and nonlinear Hall effects driven by the BCD are completely suppressed in an ideal centrosymmetric bilayer. In contrast, the quantum metric is even under both $\mathcal{P}$ and $\mathcal{T}$, satisfying $g^{ab}_n(\mathbf{k})=g^{ab}_n(-\mathbf{k})$. Its contribution to the nonlinear Hall response can be written as \cite{Chong_Wang-2021-PRL-CuMnAs,Huiying_Liu-2021-PRL-Mn2Au,Gong-2024-QMD-theory}:

\begin{equation}
\chi^\mathrm{QMD}_{abc}
= e^2 \sum_{l}
\left(v_l^a \mathcal{G}_l^{bc}-v_l^b \mathcal{G}_l^{ac}\right)\,
\partial_{\varepsilon_l} f_l ,
\end{equation}
where $l\equiv(n,\mathbf{k})$ labels a Bloch state with band index $n$ and wave vector $\mathbf{k}$, and $a,b,c$ are Cartesian indices. The group velocity is $v_l^a=\frac{1}{\hbar}\partial^{a}_\mathbf{k}\varepsilon_l$, where $\varepsilon_l$ is the eigenenergy and $\partial^b_\mathbf{k}\equiv\partial/\partial k_b$.
Here, the $\mathcal{G}_l^{ab}\equiv2e\sum_{\varepsilon_n\neq\varepsilon_m}\mathcal{R}^{nm}_{a,\mathbf{k}}\mathcal{R}^{mn}_{b,\mathbf{k}}/(\varepsilon_\mathbf{k}^n-\varepsilon_\mathbf{k}^m)$ is the Berry connection polarizability (BCP), which includes the band resolved quantum metric ($g_l^{ab}\equiv \sum_{\varepsilon_n\neq\varepsilon_m}\mathcal{R}^{nm}_{a,\mathbf{k}}\mathcal{R}^{mn}_{b,\mathbf{k}}$). The Berry connection is $\mathcal{R}^{nm}_{a,\mathbf{k}}\equiv i \langle u_n \,|\, \partial^{a} _\mathbf{k} u_m \rangle$ and $\partial_{\varepsilon_l} f_l$ indicates that the response is dominated by states near the Fermi level, $\partial_{\varepsilon_l}\equiv\partial/\partial \varepsilon_l$. Crucially, because $v_l^a$ is odd under $\mathcal{P}$ and $\mathcal{T}$ while $\mathcal{G}_l^{ab}$ is even, the integrand is odd in $k$, ensuring that the net charge response $\chi^{\mathrm{QMD}}_{abc}$ vanishes upon BZ integration in a centrosymmetric bilayer. However, the integrand can be locally finite and valley-contrasting: contributions from time-reversal-related valleys ($K$ and $K'$)  are equal in magnitude but opposite in sign. This symmetry structure makes valley-resolved detection, such as the NVHE, a clean and symmetry-preserving route to probe quantum metric physics, even when the integrated charge response is zero. Notably, a remaining threefold rotation $C_{3z}$ would otherwise prohibit the Hall-type second-order tensor components, including those enabling valley-contrasting responses. Therefore, breaking $C_{3z}$ via uniaxial strain is essential to release a finite NVHE response. For Hall-type in-plane nonlinear transport driven by the QMD, the relevant nonvanishing tensor components reduce to two candidates: $\chi^{\mathrm{QMD}}_{xyy}$ and $\chi^{\mathrm{QMD}}_{yxx}$. When strain is applied along the zigzag ($x$) direction, the residual symmetries (e.g., $C_{2x}$ and $M_x$), allow a nonzero $\chi^{\mathrm{QMD}}_{xyy}$ while forbidding $\chi^{\mathrm{QMD}}_{yxx}$, which provides clear guidance for the experimental configuration \cite{Zhichao_Zhou-2025-PRB-NVHE-2L-MoS2}.

In contrast to the centrosymmetric 2H stacking, bilayer MoS$_2$ can also adopt a noncentrosymmetric 3R stacking (see Fig. S1) \cite{Park-2019-Scientific_reports-3R_MoS2, supp}. The structure transition from 2H to 3R-type stacking removes the $180^\circ$ layer rotation characteristic of the 2H phase, thereby breaking inversion symmetry and permitting a finite out-of-plane ferroelectric (FE) polarization, arising from the non-cancellation of layer dipoles. The AB- and BA-stacked registries correspond to opposite 3R-type arrangements and therefore exhibit opposite polarization. Crucially, these slide-switchable FE states are experimentally reversible via interlayer sliding, establishing bilayer 3R-MoS$_2$ as a robust, electrically tunable platform for controlling polarization in vdW systems \cite{Meng_Peng-2022-sliding-MoS2,Fan_Aiqing-2025-Science_advance-sliding_MoS2}.

Building on these insights, we construct a four-layer MoS$_2$ system that integrates the robust centrosymmetry of a 2H-stacked core with the switchable polarization of 3R ferroelectric bilayers. As illustrated in Fig. 1, the inner two layers adopt a conventional 2H arrangement, ensuring global inversion symmetry. The outer bilayers are arranged in 3R-like AB or BA configurations, each possessing an out-of-plane ferroelectric polarization. Crucially, their polarizations are oriented in opposite directions, creating an antiferroelectric (AFE) motif, ABBA or BAAB, where the layer dipoles cancel, preserving overall centrosymmetry. Remarkably, the ABBA and BAAB configurations are interconvertible via interlayer sliding of the outer layers, enabling reversible AFE switching without breaking global inversion or introducing magnetism. When combined with uniaxial strain to break $C_{3z}$ symmetry, this architecture provides a symmetry-protected, electrically tunable platform for realizing and controlling a QMD-driven NVHE. A comprehensive summary of the symmetry conditions and accessible nonlinear responses for various MoS$_2$ stackings, including monolayer (1L), bilayer (2L-2H and 2L-3R), and our four-layer AFE (4L-AFE) design, is presented in Table I.

We next examine the energetics of AFE switching between ABBA and BAAB stackings. Figure 1(a,b) illustrates two representative sliding pathways: Path 1 involves sequential lateral sliding of the top and then the bottom layer, while Path 2 entails simultaneous sliding of both outer layers. In both cases, the inner bilayer retains the 2H stacking, ensuring that global inversion symmetry is restored in the final states. The corresponding energy profiles are shown in Fig. 1(c).
{\color{black}The two end states, ABBA and BAAB, are found to be nearly degenerate in energy, consistent with the bistable nature of the AFE degree of freedom, as discussed in the Supplementary Material}\cite{supp}.
The switching barrier for Path 1 is approximately 17.7 meV per unit cell (meV/u.c.), which is roughly half that of Path 2 (35.5 meV/u.c.).
These low energy barriers confirm that AFE switching is energetically feasible, with Path 1 being the preferential pathway under realistic experimental conditions. The small energy scales further imply that switching can be driven by moderate in-plane electric fields or mechanical shear, offering a practical handle for device implementation.

{\color{black} A defining feature of our four-layer design is the distinct layer-distribution character of the conduction-band states near the $K$ and $K'$ valleys. As shown in Figs.~2(a) and 2(b), we characterize their layer distribution through a continuous expectation value
\begin{equation}
\langle |z| \rangle = \frac{\sum_i w_i |z_i-z_c|}{\sum_i w_i},
\end{equation}
where $w_i$ is the projected weight on atomic site $i$, $z_i$ is the out-of-plane coordinate of that site, and $z_c$ denotes the center of the multilayer. In this representation, smaller values of $\langle |z| \rangle$ indicate that the state is distributed closer to the inner layers, whereas larger values indicate stronger outer-layer character. The results show that the states near the conduction-band minimum in ABBA are characterized by larger $\langle |z| \rangle$, while those in BAAB exhibit smaller $\langle |z| \rangle$. This layer reversal is a direct manifestation of the opposite antiferroelectric polarization patterns in the two stackings. Consequently, the valley-resolved QMD inherits this distinct layer-distribution tendency: in ABBA the dominant QMD contributions are associated with states of stronger outer-layer character [Figs.~2(c,d)], whereas in BAAB they are associated with states of stronger inner-layer character [Figs.~2(e,f)]. }
This layer–valley–QMD locking establishes a clear microscopic link between stacking order, band geometry, and the resulting nonlinear valley Hall response, enabling tunable control of quantum-metric transport through interlayer sliding.

To quantitatively compare the two stacking configurations, Figs. 3(a,b) present the calculated nonlinear valley Hall conductivity (NVHC) for ABBA and BAAB. The results reveal a pronounced layer-selective contrast: the NVHC contribution from the outer layers is substantially larger than that from the inner layers. Accordingly, ABBA, whose CBM states are primarily localized on the outer layers, exhibits a stronger NVHC than BAAB, where the relevant states are concentrated on the inner layers. This contrast is rooted in the quantum geometry of the conduction bands. The BCP of band $n$, introduced in Eq. (1), can be written explicitly as
\begin{equation}
\mathcal{G}^{ab}_n(\mathbf{k}) = 2\sum_{m \neq n} 
\frac{\mathrm{Re}\left[\langle u_n | \partial^a_\mathbf{k} \hat{H} | u_m \rangle 
\langle u_m | \partial^b_\mathbf{k} \hat{H} | u_n \rangle \right]}
{(\varepsilon_m - \varepsilon_n)^3}.
\end{equation}
Here, $\hat{H}$ is the Hamiltonian and $| u_n \rangle$ denotes the Bloch eigenstate. The cubic energy denominator implies that band crossings or near-degeneracies near the CBM can produce sharply enhanced quantum-metric ``hot spots'' \cite{Kam_Das-2025-Nanoletter-QMD_CrSBr}. In our four-layer MoS$_2$, such features are more pronounced for the outer-layer conduction states around the $K$ and $K'$ valleys, leading to larger quantum-metric peaks and thus a stronger QMD-driven NVHC from the outer layers (see Fig.~S4 \cite{supp}). This direct connection between layer-selective band geometry and layer-selective nonlinear response illustrates how stacking-controlled quantum geometry can be harnessed to control valley transport.

\begin{figure}
\includegraphics[scale=0.22]{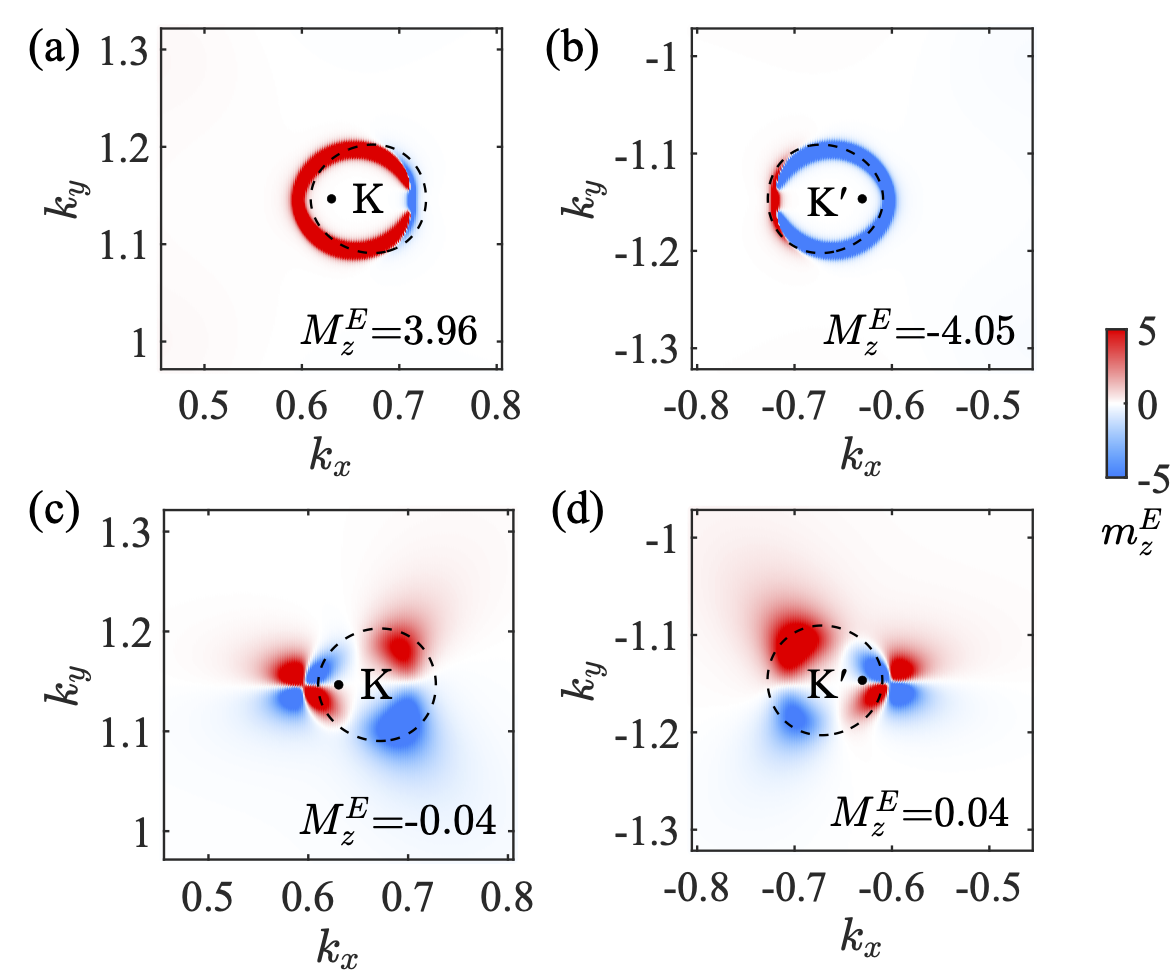}
\caption{
Momentum-space distribution of the EOMM, $m_z^{E}(\mathbf{k})$, for the lowest conduction band of strained four-layer MoS$_2$ under an in-plane electric field $E_y=1~\mathrm{V/\mu m}$. Panels (a,b) and (c,d) show $m_z^{E}(\mathbf{k})$ near the $K$ and $K'$ valleys for ABBA and BAAB stacking configurations, respectively. The color scale denotes $m_z^{E}(\mathbf{k})$ in units of $(e/\hbar)\,\text{\AA}^2\,\mathrm{meV}$.
The dashed contours mark the Fermi level, defining the integration region for the valley-resolved field-induced orbital magnetization $M^{E}$ (units of $(e/\hbar)\,\mathrm{meV}$). The resulting magnetization exhibits clear valley contrasting: for ABBA (at $\mu=0.37$~eV), $M_{K}^{E}=3.96\,(e/\hbar)\,\mathrm{meV}$ and $M_{K'}^{E}=-4.05\,(e/\hbar)\,\mathrm{meV}$; for BAAB (at $\mu=0.39$~eV), $M_{K}^{E}=-0.04\,(e/\hbar)\,\mathrm{meV}$ and $M_{K'}^{E}=0.04\,(e/\hbar)\,\mathrm{meV}$. These opposite-sign, valley‑locked orbital moments provide a direct microscopic link between quantum‑metric geometry and the nonlinear valley Hall effect. }
\end{figure}

{\color{black}
To visualize the calculated layer-resolved NVHE, Figs.~3(c,d) provide schematic illustrations of its stacking-dependent layer character. Consistent with the layer-resolved band projections and QMD distributions in Fig.~2 and the valley-contrasting NVHC shown in Figs.~3(a,b), the dominant nonlinear valley Hall response evolves from predominantly outer-layer character in ABBA to predominantly inner-layer character in BAAB. In particular, the arrows in Fig.~3(c) depict the valley-contrasting transverse deflection of carriers in the dominant transport layer under an in-plane driving field, providing intuitive guidance for future experimental identification of the layer-resolved nonlinear valley Hall effect.}

Analogous to the spin Hall effect, which generates spin accumulation at sample edges, and the valley Hall effect, which produces valley-contrasting orbital moments, the NVHE creates valley-polarized carriers distinguished by their electric-field-induced orbital magnetic moment (EOMM). To elucidate the underlying quantum-geometric mechanism, we compute the EOMM near the $K$ and $K'$ valleys. For band $n$ under an electric field $E_d$ , the EOMM is expressed as:
\cite{Qian_Niu-2021-PRB-OMM,Bangguo_Xiong-2021-PRB-OMM,Kamal_Das-2024-PRL-NVHE}
\begin{equation}
m_a^E(n)
= \sum_{m \neq n}
\left[
2e\,\frac{\mathrm{Re}\big(\mathcal{M}_{mn}^a \mathcal{R}_{nm}^d\big)}{\varepsilon_n - \varepsilon_m}
+ \frac{e^2}{2\hbar}\,\epsilon_{abc}\,\partial^b_\mathbf{k} \mathcal{G}_{mn}^{cd}
\right] E_d .
\end{equation}
Here, $\mathcal{M}_{mn}^{a}$ is the interband EOMM, defined by $\mathcal{M}_{mn}^{a}
= \frac{e}{2}\sum_{l\neq n}
\epsilon_{abc}\big(v^{b}_{ml}+ v^{b}_{n}\,\delta_{ml}\big) \mathcal{R}^{c}_{lm}
$, where $v^{b}_{ml}=\langle u_m \,|\, \frac{1}{\hbar}\partial^{b}_\mathbf{k}\hat{H} \,|\, u_l \rangle$ the matrix element of the velocity operator, $v^b_n\equiv v^b_{nn}$ and
$\mathcal{R}_{lm}^{d}= i \langle u_l \,|\, \partial^{d} _\mathbf{k} u_m \rangle$
is the interband Berry connection. In addition, $\epsilon_{abc}$ is the Levi--Civita tensor, $E_d$ denotes the applied electric field (with $d = x, y, z$). Then the net field-induced orbital magnetization can be obviously defined as,
\begin{equation}
M_a^{E}(\mu)
=  \sum_{l\equiv(n,\mathbf{k})}
m_a^{E}(n)\, f_l,
\label{eq:valleyME}
\end{equation}
where $\mu$ is the chemical potential and $f_l$ is the Fermi--Dirac distribution. Notably, the Berry connection and velocity matrix can only have in-plane components for a 2D system. Therefore, with an in-plane driving field in our system, only the out-of-plane component survives, i.e., $a=z$.

Figure 4 presents the momentum-resolved EOMM distributions around the CBM for ABBA and BAAB. While the EOMM exhibits both positive and negative contributions within each valley, the valley-integrated orbital magnetization is valley-contrasting: the total EOMM at $K$ and $K'$ is equal in magnitude but opposite in sign. As a result, carriers from opposite valleys acquire opposite orbital moments under the same driving field. Because the system remains centrosymmetric and nonmagnetic, the BZ-integrated orbital magnetization vanishes by symmetry, even though the EOMM is locally finite in momentum space.  This symmetry-protected valley contrast provides a direct microscopic connection between the QMD-governed band geometry and the NVHE: the driving field induces opposite orbital responses in the two valleys, which do not sum to a net magnetization but instead manifest as a nonlinear valley Hall current, a purely geometric signature accessible in valley-resolved transport.

\section{IV. Conclusion}

In summary, we have proposed a symmetry-engineered four-layer MoS$_2$ platform with sliding-controlled antiferroelectric order to realize a quantum-metric-driven NVHE. By preserving both $\mathcal P$ and $\mathcal T$ symmetries while breaking $C_{3z}$ rotation via uniaxial strain, the system develops nonzero valley-contrasting QMDs, enabling a robust and electrically tunable NVHE. Unlike previously studied bilayer systems, our design incorporates reversible ABBA $\leftrightarrow$BAAB interlayer sliding, which inverts the layer-polarization pattern without breaking global inversion, thereby adding an active nonvolatile control knob. First-principles and Wannier-based tight-binding calculations demonstrate that the NVHE is strongly layer-resolved and outer-layer–dominated. Band crossings near the CBM at $K$ and $K'$ produce sharply enhanced, valley-contrasting QMD ``hot spots'' on specific layers, leading to a pronounced QMD-driven nonlinear valley Hall conductivity. This quantum-geometric structure also gives rise to a finite, valley-contrasting electric-field-induced orbital magnetic moment, offering a clear microscopic picture in which an applied field drives opposite orbital currents in the two valleys while preserving zero net magnetization. Our work establishes a practical and versatile pathway to activate and control hidden quantum-geometric responses in centrosymmetric, nonmagnetic 2D materials, and demonstrates how symmetry engineering, stacking order, and sliding ferroelectricity can be combined to design tunable valleytronic and quantum-information–oriented devices.

\section{Acknowledgements}
This work was supported by the National Key R \& D Program of China (nos. 2022YFA1402903 and 2024YFA1409101), the National Natural Science Foundation of China under Grant No. 12374059, and the Shenzhen Science and Technology Program (Grant No. JCYJ20240813095301003).  Part of this work was supported by the Quantum Science Center of Guangdong-Hong Kong-Macao Greater Bay Area (Guangdong). Computational time was supported by the Center for Computational Science and Engineering of Southern University of Science and Technology.

\bibliographystyle{apsrev4-1-etal-title_10authors}
\bibliography{main}

@article{Di_Xiao-2007-PRL-valley-grahene,
  title = {Valley-Contrasting Physics in Graphene: Magnetic Moment and Topological Transport},
  author = {Xiao, Di and Yao, Wang and Niu, Qian},
  journal = {Phys. Rev. Lett.},
  volume = {99},
  issue = {23},
  pages = {236809},
  numpages = {4},
  year = {2007},
  month = {Dec},
  publisher = {American Physical Society},
  doi = {10.1103/PhysRevLett.99.236809},
  url = {https://link.aps.org/doi/10.1103/PhysRevLett.99.236809}
}

@article{Di_Xiao-2012-PRL-MoS2-valley,
  title = {Coupled Spin and Valley Physics in Monolayers of ${\mathrm{MoS}}_{2}$ and Other Group-VI Dichalcogenides},
  author = {Xiao, Di and Liu, Gui-Bin and Feng, Wanxiang and Xu, Xiaodong and Yao, Wang},
  journal = {Phys. Rev. Lett.},
  volume = {108},
  issue = {19},
  pages = {196802},
  numpages = {5},
  year = {2012},
  month = {May},
  publisher = {American Physical Society},
  doi = {10.1103/PhysRevLett.108.196802},
  url = {https://link.aps.org/doi/10.1103/PhysRevLett.108.196802}
}

@article{Park-2019-Scientific_reports-3R_MoS2,
  title={Ferroelectric switching in bilayer {3R} {MoS$_2$} via interlayer shear mode driven by nonlinear phononics},
  author={Park, Jaehong and Yeu, In Won and Han, Gyuseung and Hwang, Cheol Seong and Choi, Jung-Hae},
  journal={Sci. Rep.},
  volume={9},
  number={1},
  pages={14919},
  year={2019},
  publisher={Nature Publishing Group UK London},
  url = {https://www.nature.com/articles/s41598-019-50293-y}
}

@article{Meng_Peng-2022-sliding-MoS2,
  title={Sliding induced multiple polarization states in two-dimensional ferroelectrics},
  author={Meng, Peng and Wu, Yaze and Bian, Renji and Pan, Er and Dong, Biao and Zhao, Xiaoxu and Chen, Jiangang and Wu, Lishu and Sun, Yuqi and Fu, Qundong and others},
  journal={Nat. Commun.},
  volume={13},
  number={1},
  pages={7696},
  year={2022},
  publisher={Nature Publishing Group UK London},
  url = {https://www.nature.com/articles/s41467-022-35339-6}
}

@article{Cao_Jin-2025-Nonlocal-NVHE,
  title = {Nonlocal transport from nonlinear valley responses},
  author = {Cao, Jin and Wang, Hui and Lai, Shen and Xiao, Cong and Yang, Shengyuan A.},
  journal = {Phys. Rev. B},
  volume = {112},
  issue = {12},
  pages = {L121404},
  numpages = {6},
  year = {2025},
  month = {Sep},
  publisher = {American Physical Society},
  doi = {10.1103/pjmx-rmnt},
  url = {https://link.aps.org/doi/10.1103/pjmx-rmnt}
}

@article{Fan_Aiqing-2025-Science_advance-sliding_MoS2,
  title={Tailored sliding ferroelectricity for ultrahigh fatigue resistance in stacked trilayer {MoS$_2$} crystals},
  author={Fan, Aiqing and Zhang, Qing and Yang, Zihao and Li, Lin and Li, Menghan and Zhang, Keyu and Gao, Junfeng and Wu, Fan and Wu, Menghao and Geng, Dechao and others},
  journal={Sci. Adv.},
  volume={11},
  number={31},
  pages={eadx8192},
  year={2025},
  publisher={American Association for the Advancement of Science},
  url = {https://www.science.org/doi/abs/10.1126/sciadv.adx8192}
}

@article{xiong-2025-JPCM-MoS2_BCD,
  title={Strain tuning of the nonlinear anomalous {Hall} effect in {MoS$_2$} monolayer},
  author={Xiong, Yuebei and Gong, Zhirui and Jin, Hao},
  journal={J. Phys. Condens. Matter},
  volume={37},
  number={19},
  pages={195301},
  year={2025},
  publisher={IOP Publishing},
  url = {https://iopscience.iop.org/article/10.1088/1361-648X/adc64a/meta}
}

@article{You_Jhih-Shih-2018-PRB-MoS2-BCP,
  title = {Berry curvature dipole current in the transition metal dichalcogenides family},
  author = {You, Jhih-Shih and Fang, Shiang and Xu, Su-Yang and Kaxiras, Efthimios and Low, Tony},
  journal = {Phys. Rev. B},
  volume = {98},
  issue = {12},
  pages = {121109},
  numpages = {6},
  year = {2018},
  month = {Sep},
  publisher = {American Physical Society},
  doi = {10.1103/PhysRevB.98.121109},
  url = {https://link.aps.org/doi/10.1103/PhysRevB.98.121109}
}

@article{Liang_Fu-2015-PRL-BCD,
  title = {Quantum Nonlinear {Hall} Effect Induced by Berry Curvature Dipole in Time-Reversal Invariant Materials},
  author = {Sodemann, Inti and Fu, Liang},
  journal = {Phys. Rev. Lett.},
  volume = {115},
  issue = {21},
  pages = {216806},
  numpages = {5},
  year = {2015},
  month = {Nov},
  publisher = {American Physical Society},
  doi = {10.1103/PhysRevLett.115.216806},
  url = {https://link.aps.org/doi/10.1103/PhysRevLett.115.216806}
}

@article{Kamal_Das-2024-PRL-NVHE,
  title={Nonlinear valley {Hall} effect},
  author={Das, Kamal and Ghorai, Koushik and Culcer, Dimitrie and Agarwal, Amit},
  journal={Phys. Rev. Lett.},
  volume={132},
  number={9},
  pages={096302},
  year={2024},
  publisher={APS},
  url = {https://journals.aps.org/prl/abstract/10.1103/PhysRevLett.132.096302}
}

@article{Zhichao_Zhou-2025-PRB-NVHE-2L-MoS2,
  title={Nonlinear valley Hall effect in a bilayer transition metal dichalcogenide},
  author={Zhou, Zhichao and Fang, Ruijing and Zhang, Zhen and Wang, Xiaoyu and Rong, Jiayan and Li, Xiao},
  journal={Phys. Rev. B},
  volume={112},
  number={4},
  pages={045432},
  year={2025},
  publisher={APS},
  url ={https://journals.aps.org/prb/abstract/10.1103/fs39-vgq6}
}

@article{Huangling_Zeng-2012-Nat_Nanotech-Valley-MoS2-experiment-optical,
  title={Valley polarization in {MoS$_2$} monolayers by optical pumping},
  author={Zeng, Hualing and Dai, Junfeng and Yao, Wang and Xiao, Di and Cui, Xiaodong},
  journal={Nat. Nanotechnol.},
  volume={7},
  number={8},
  pages={490--493},
  year={2012},
  publisher={Nature Publishing Group UK London},
  url = {https://www.nature.com/articles/nnano.2012.95}
}

@article{Mengqiao_Sui-2015-Nat_Phys-VHE-2L-graphen-experiment,
  title={Gate-tunable topological valley transport in bilayer graphene},
  author={Sui, Mengqiao and Chen, Guorui and Ma, Liguo and Shan, Wen-Yu and Tian, Dai and Watanabe, Kenji and Taniguchi, Takashi and Jin, Xiaofeng and Yao, Wang and Xiao, Di and others},
  journal={Nat. Phys.},
  volume={11},
  number={12},
  pages={1027--1031},
  year={2015},
  publisher={Nature Publishing Group UK London},
  url = {https://www.nature.com/articles/nphys3485}
}

@article{Y_Shimazaki-2015-Nat_Phys-VHE-2L-graphen-experiment,
  title={Generation and detection of pure valley current by electrically induced Berry curvature in bilayer graphene},
  author={Shimazaki, Yuya and Yamamoto, Michihisa and Borzenets, Ivan V and Watanabe, Kenji and Taniguchi, Takashi and Tarucha, Seigo},
  journal={Nat. Phys.},
  volume={11},
  number={12},
  pages={1032--1036},
  year={2015},
  publisher={Nature Publishing Group UK London},
  url = {https://www.nature.com/articles/nphys3551}
}

@article{Chong_Wang-2021-PRL-CuMnAs,
  title = {Intrinsic Nonlinear {Hall} Effect in Antiferromagnetic Tetragonal {CuMnAs}},
  author = {Wang, Chong and Gao, Yang and Xiao, Di},
  journal = {Phys. Rev. Lett.},
  volume = {127},
  issue = {27},
  pages = {277201},
  numpages = {6},
  year = {2021},
  month = {Dec},
  publisher = {American Physical Society},
  doi = {10.1103/PhysRevLett.127.277201},
  url = {https://link.aps.org/doi/10.1103/PhysRevLett.127.277201}
}

@article{Huiying_Liu-2021-PRL-Mn2Au,
  title = {Intrinsic Second-Order Anomalous Hall Effect and Its Application in Compensated Antiferromagnets},
  author = {Liu, Huiying and Zhao, Jianzhou and Huang, Yue-Xin and Wu, Weikang and Sheng, Xian-Lei and Xiao, Cong and Yang, Shengyuan A.},
  journal = {Phys. Rev. Lett.},
  volume = {127},
  issue = {27},
  pages = {277202},
  numpages = {6},
  year = {2021},
  month = {Dec},
  publisher = {American Physical Society},
  doi = {10.1103/PhysRevLett.127.277202},
  url = {https://link.aps.org/doi/10.1103/PhysRevLett.127.277202}
}

@article{Anyuan_Gao-2023-Science-QMD-MnBiTe-experiment,
  title={Quantum metric nonlinear {Hall} effect in a topological antiferromagnetic heterostructure},
  author={Gao, Anyuan and Liu, Yu-Fei and Qiu, Jian-Xiang and Ghosh, Barun and V. Trevisan, Tha{\'\i}s and Onishi, Yugo and Hu, Chaowei and Qian, Tiema and Tien, Hung-Ju and Chen, Shao-Wen and others},
  journal={Science},
  volume={381},
  number={6654},
  pages={181--186},
  year={2023},
  publisher={American Association for the Advancement of Science},
  url = {https://www.science.org/doi/abs/10.1126/science.adf1506}
}

@article{Naizhou_Wang-2023-Nature-QMD-experiment,
  title={Quantum-metric-induced nonlinear transport in a topological antiferromagnet},
  author={Wang, Naizhou and Kaplan, Daniel and Zhang, Zhaowei and Holder, Tobias and Cao, Ning and Wang, Aifeng and Zhou, Xiaoyuan and Zhou, Feifei and Jiang, Zhengzhi and Zhang, Chusheng and others},
  journal={Nature},
  volume={621},
  number={7979},
  pages={487--492},
  year={2023},
  publisher={Nature Publishing Group UK London},
  url = {https://www.nature.com/articles/s41586-023-06363-3}
}

@article{Gong-2024-QMD-theory,
  title         = {Scaling analysis of quantum geometry in second-order nonlinear transport},
  author        = {Gong, Zhen-Hao and Du, Z.-Z. and Sun, Hai-Peng and Lu, Hai-Zhou and Xie, X. C.},
  journal       = {Sci. Bull.},
  volume        = {71},
  number        = {17},
  pages         = {4422--4428},
  year          = {2026},
  doi           = {10.1016/j.scib.2026.07.062},
  eprint        = {2410.04995},
  archivePrefix = {arXiv},
  primaryClass  = {cond-mat.mes-hall}
}

@article{ZZ_Du-2021-NC-NHE,
  title={Quantum theory of the nonlinear {Hall} effect},
  author={Du, ZZ and Wang, CM and Sun, Hai-Peng and Lu, Hai-Zhou and Xie, XC},
  journal={Nat. Commun.},
  volume={12},
  number={1},
  pages={5038},
  year={2021},
  publisher={Nature Publishing Group UK London},
  url = {https://www.nature.com/articles/s41467-021-25273-4}
}

@misc{supp,
  author       = {Jiang, Wen-Xin and Gong, Zhen-Hao and Gui, Zhigang and Farooq, M. Umar and Huang, Li},
  title        = {{Supporting information for `Tunable Quantum-Metric-Driven Nonlinear Valley Hall Effect in \(\mathcal{P,T}\)-symmetric Sliding Antiferroelectric Four-Layer MoS$_2$'}},
  year         = {2026},
  howpublished = {Supplemental Material accompanying \href{https://doi.org/10.1103/p9sb-k9fj}{Phys. Rev. B \textbf{113}, 235116}},
  note         = {Includes additional structural, electronic, and transport data supporting the main text. It contains schematic illustrations of different stacking configurations and symmetry properties of bilayer and multilayer {MoS$_2$}, layer-resolved and orbital-resolved electronic band structures under uniaxial strain, quantum-metric-dipole (QMD)--projected band structures around the $K$ and $K'$ valleys, and detailed Fermi-energy-dependent nonlinear valley {Hall} conductivity $\chi^{\mathrm{QMD}}_{xyy}$ for ABBA and BAAB stacking configurations under varying strain strengths. These results provide microscopic insight into the layer selectivity, orbital character, and strain-tunable quantum-geometric origin of the nonlinear valley {Hall} effect, and further substantiate the symmetry-based mechanism and sliding-controlled tunability discussed in the main text.}
}

@article{JP_Provost-1980-Riemannian,
  title={Riemannian structure on manifolds of quantum states},
  author={Provost, JP and Vallee, G},
  journal={ Commun. Math. Phys.},
  volume={76},
  number={3},
  pages={289--301},
  year={1980},
  publisher={Springer},
  url = {https://link.springer.com/article/10.1007/bf02193559}
}

@article{Resta-2011-geometrical_theory,
  title={The insulating state of matter: a geometrical theory},
  author={Resta, Raffaele},
  journal={Eur. Phys. J. B},
  volume={79},
  number={2},
  pages={121--137},
  year={2011},
  publisher={Springer},
  url = {https://link.springer.com/article/10.1140/epjb/e2010-10874-4}
}

@article{Rahul_Roy-2014-Phys_Rev_B-band_geometry,
  title={Band geometry of fractional topological insulators},
  author={Roy, Rahul},
  journal={Phys. Rev. B},
  volume={90},
  number={16},
  pages={165139},
  year={2014},
  publisher={APS},
  url = {https://journals.aps.org/prb/abstract/10.1103/PhysRevB.90.165139}
}

@article{Peotta_Sebastiano-2015-Nat_Commun-superfluidity_topology,
  title={Superfluidity in topologically nontrivial flat bands},
  author={Peotta, Sebastiano and T{\"o}rm{\"a}, P{\"a}ivi},
  journal={Nat. Commun.},
  volume={6},
  number={1},
  pages={8944},
  year={2015},
  publisher={Nature Publishing Group UK London},
  url = {https://www.nature.com/articles/ncomms9944}
}

@article{Torma-2022-Nat_Rev_Phys-superconductivity_superfluidity_quantum_geometry,
  title={Superconductivity, superfluidity and quantum geometry in twisted multilayer systems},
  author={T{\"o}rm{\"a}, P{\"a}ivi and Peotta, Sebastiano and Bernevig, Bogdan A},
  journal={Nat. Rev. Phys.},
  volume={4},
  number={8},
  pages={528--542},
  year={2022},
  publisher={Nature Publishing Group UK London},
  url = {https://www.nature.com/articles/s42254-022-00466-y}
}

@article{Torma-2023-PRL-Essay_QMD,
  title={Essay: Where can quantum geometry lead us?},
  author={T{\"o}rm{\"a}, P{\"a}ivi},
  journal={Phys. Rev. A},
  volume={131},
  number={24},
  pages={240001},
  year={2023},
  publisher={APS},
  url = {https://journals.aps.org/prl/abstract/10.1103/PhysRevLett.131.240001}
}

@article{Hetenyi-2023-PRA-geometry_qauntum_state,
  title={Fluctuations, uncertainty relations, and the geometry of quantum state manifolds},
  author={Het{\'e}nyi, Bal{\'a}zs and L{\'e}vay, P{\'e}ter},
  journal={Phys. Rev. A},
  volume={108},
  number={3},
  pages={032218},
  year={2023},
  publisher={APS},
  url = {https://journals.aps.org/pra/abstract/10.1103/PhysRevA.108.032218}
}

@article{Reimers-2023-NC-current-domain-Mn2Au,
  title={Current-driven writing process in antiferromagnetic {Mn$_2$Au} for memory applications},
  author={Reimers, Sonka and Lytvynenko, Yaryna and Niu, YR and Golias, Evangelos and Sarpi, Brice and Veiga, LSI and Denneulin, Thibauld and Kovacs, Andras and Dunin-Borkowski, RE and Bl{\"a}{\ss}er, Jonathan and others},
  journal={Nat. Commun.},
  volume={14},
  number={1},
  pages={1861},
  year={2023},
  publisher={Nature Publishing Group UK London},
  url = {https://www.nature.com/articles/s41467-023-37569-8}
}

@article{Sheng_Peng-2022-thermal_contribution,
  author  = {Sheng, Peng and Zhao, Zhengyang and Benally, Onri Jay and Zhang, Delin and Wang, Jian-Ping},
  title   = {Thermal contribution in the electrical switching experiments with heavy metal/antiferromagnet structures},
  journal = {J. Appl. Phys.},
  year    = {2022},
  volume  = {132},
  number  = {7},
  pages   = {073902},
  doi     = {10.1063/5.0098631},
  url = {https://pubs.aip.org/aip/jap/article/132/7/073902/2837324}
}

@article{Qian_Niu-2021-PRB-OMM,
  title={Thermoelectric generation of orbital magnetization in metals},
  author={Xiao, Cong and Liu, Huiying and Zhao, Jianzhou and Yang, Shengyuan A and Niu, Qian},
  journal={Phys. Rev. B},
  volume={103},
  number={4},
  pages={045401},
  year={2021},
  publisher={APS},
  url = {https://journals.aps.org/prb/abstract/10.1103/PhysRevB.103.045401}
}

@article{Bangguo_Xiong-2021-PRB-OMM,
  title={Adiabatically induced orbital magnetization},
  author={Xiao, Cong and Ren, Yafei and Xiong, Bangguo},
  journal={Phys. Rev. B},
  volume={103},
  number={11},
  pages={115432},
  year={2021},
  publisher={APS},
  url = {https://link.aps.org/doi/10.1103/PhysRevB.103.115432}
}

@article{Kam_Das-2025-Nanoletter-QMD_CrSBr,
author = {Das, Kamal and Zhao, Yufei and Yan, Binghai},
title = {Surface-Dominated Quantum-Metric-Induced Nonlinear Transport in the Layered Antiferromagnet {CrSBr}},
journal = {Nano Lett.},
volume = {25},
number = {23},
pages = {9189-9196},
year = {2025},
doi = {10.1021/acs.nanolett.5c00195},
    note ={PMID: 40424510},

URL = { 
    
        https://doi.org/10.1021/acs.nanolett.5c00195
    
    

}

}

@article{kresse-1994-GGA,
  title={Theory of the crystal structures of selenium and tellurium: the effect of generalized-gradient corrections to the local-density approximation},
  author={Kresse, G and Furthm{\"u}ller, J and Hafner, JJPRB},
  journal={Phys. Rev. B},
  volume={50},
  number={18},
  pages={13181},
  year={1994},
  publisher={APS},
  url = {https://journals.aps.org/prb/abstract/10.1103/PhysRevB.50.13181}
}

@article{kresse-1996-efficient-iterative-schemes,
  title={Efficient iterative schemes for ab initio total-energy calculations using a plane-wave basis set},
  author={Kresse, Georg and Furthm{\"u}ller, J{\"u}rgen},
  journal={Phys. Rev. B},
  volume={54},
  number={16},
  pages={11169},
  year={1996},
  publisher={APS},
  url = {https://journals.aps.org/prb/abstract/10.1103/PhysRevB.54.11169}
}

@article{perdew-1996-PBE,
  title={Generalized gradient approximation made simple},
  author={Perdew, John P and Burke, Kieron and Ernzerhof, Matthias},
  journal={Phys. Rev. Lett.},
  volume={77},
  number={18},
  pages={3865},
  year={1996},
  publisher={APS},
  url = {https://journals.aps.org/prl/abstract/10.1103/PhysRevLett.77.3865}
}

@article{kresse-1999-PAW,
  title={From ultrasoft pseudopotentials to the projector augmented-wave method},
  author={Kresse, Georg and Joubert, Daniel},
  journal={Phys. Rev. B},
  volume={59},
  number={3},
  pages={1758},
  year={1999},
  publisher={APS},
  url = {https://journals.aps.org/prb/abstract/10.1103/PhysRevB.59.1758}
}

@article{Blochl-1994-PAW,
  title={Projector augmented-wave method},
  author={Bl{\"o}chl, Peter E},
  journal={Phys. Rev. B},
  volume={50},
  number={24},
  pages={17953},
  year={1994},
  publisher={APS},
  url = {https://journals.aps.org/prb/abstract/10.1103/PhysRevB.50.17953}
}

@article{Grimme-2011-DFT_D3-JCC,
  title={Effect of the damping function in dispersion corrected density functional theory},
  author={Grimme, Stefan and Ehrlich, Stephan and Goerigk, Lars},
  journal={J. Comput. Chem.},
  volume={32},
  number={7},
  pages={1456--1465},
  year={2011},
  publisher={Wiley Online Library},
  url = {https://onlinelibrary.wiley.com/doi/abs/10.1002/jcc.21759}
}

@article{Wannier90,
title = "An updated version of wannier90: A tool for obtaining maximally-localised Wannier functions",
journal = "Comput. Phys. Commun.",
volume = "185",
number = "8",
pages = "2309 - 2310",
year = "2014",
issn = "0010-4655",
doi = "https://doi.org/10.1016/j.cpc.2014.05.003",
url = "http://www.sciencedirect.com/science/article/pii/S001046551400157X",
author = "Arash A. Mostofi and Jonathan R. Yates and Giovanni Pizzi and Young-Su Lee and Ivo Souza and David Vanderbilt and Nicola Marzari"
}

\end{document}